\documentclass[useAMS,usenatbib,usegraphicx]{mn2e}
\usepackage{bm}
\usepackage{times}

\title[Dust formation in high redshift galaxies]
{Constraining dust formation in high-redshift young galaxies}
\author[Hirashita et al.]{Hiroyuki Hirashita,$^1$\thanks{E-mail:
    hirashita@asiaa.sinica.edu.tw}
Andrea Ferrara,$^2$ Pratika Dayal$^3$ and
Masami Ouchi$^4$\\
$^1$Institute of Astronomy and Astrophysics, Academia Sinica,
P.O. Box 23-141, Taipei 10617, Taiwan\\
$^2$Scuola Normale Superiore, Piazza dei Cavalieri 7, I-56126
Pisa, Italy\\
$^3$SUPA\thanks{Scottish Universities Physics Alliance}, Institute for Astronomy, University of Edinburgh, Royal Observatory, Edinburgh, EH9 3HJ, UK\\
$^4$Institute for Cosmic Ray Research, The University of Tokyo,
Kashiwa-no-ha, Kashiwa 277-8582, Japan
}
\date{2014 June 25}

\pagerange{\pageref{firstpage}--\pageref{lastpage}} \pubyear{2014}

\begin{document}
\label{firstpage}
\maketitle

\begin{abstract}
Core-collapse supernovae (SNe) are believed to be the first significant
source of dust in the Universe. Such SNe are expected to be
the main dust producers in young high-redshift Lyman $\alpha$ emitters (LAEs)
given their young ages, providing an excellent testbed of SN dust formation models
during the early stages of galaxy evolution. We focus on the dust enrichment of
a specific, luminous LAE (Himiko, $z \simeq 6.6$) for which a stringent upper limit
of $52.1~\mu$Jy ($3\sigma$) has recently been obtained from
ALMA continuum observations at 1.2 mm.
We predict its submillimetre dust emission
using detailed models that follow SN dust enrichment and destruction
{and the equilibrium dust temperature}, and obtain
{a plausible upper limit} to the dust mass produced by a single SN:  
$m_\mathrm{d,SN} < 0.15$--$0.45~\mathrm{M}_{\sun}$, depending on the adopted
dust optical properties. These upper limits are smaller than the
dust mass deduced for SN 1987A and that predicted by dust condensation
theories, implying that dust produced in SNe are likely to be subject to reverse shock
destruction before being injected into the interstellar medium.
Finally, we provide a recipe for deriving $m_\mathrm{d,SN}$ from submillimetre observations of young, metal poor 
objects wherein condensation in SN ejecta is the dominant dust formation channel.
\end{abstract}

\begin{keywords}
dust, extinction ---
galaxies: evolution --- galaxies: high-redshift ---
galaxies: ISM --- ISM: supernova remnants ---
submillimetre: galaxies
\end{keywords}

\section{Introduction}

Dust plays a key role in determining the formation and visibility of galaxies and
their stars. Dust {surfaces are} the main site for the formation of some
molecular species, especially H$_2$
\citep[e.g.][]{gould63,cazaux04}, inducing the formation of
molecular clouds in galaxies \citep{hirashita02,yamasawa11}.
Dust cooling induces fragmentation of molecular clouds
\citep{omukai00,omukai05} and shapes the stellar initial mass function 
(IMF; \citealt{schneider06}). Moreover, dust grains absorb optical
and ultraviolet (UV) light and reprocess it into infrared bands, 
dramatically affecting the observed galaxy spectra
\citep[e.g.][]{takeuchi05}.

Over the last {decade}, pieces of evidence have accumulated
for astonishingly large ($\simeq 10^8$ M$_{\sun}$) amounts of dust
in some $z\simeq 6$ quasars (QSOs; e.g.\
\citealt{bertoldi03,priddey03,robson04,beelen06,michalowski10a}).
As QSOs are highly biased and largely evolved systems, 
systematically characterized by extreme star formation rates and
high metallicities ($\ga 1$ Z$_{\sun}$; \citealt{freudling03,juarez09}),
they cannot be 
readily used to study the initial dust production phases and sources
\citep{valiante11,kuo12}. High-redshift submillimetre-selected galaxies
(SMGs) also have a large amount of dust
by definition \citep[e.g.][and references therein]{michalowski10b}; however, they
are also evolved objects in terms of dust enrichment.
Therefore, it is 
necessary to turn our attention to younger, less evolved systems such as
Lyman $\alpha$ emitters (LAEs) to make progress on the first dust enrichment in the Universe.

While stellar evolution arguments suggest the most likely
first cosmic dust factories were core-collapse supernovae (SNe;
\citealt{todini01,nozawa07,bianchi07}), the dust mass formed per supernova remains 
a highly uncertain quantity; for reasons of brevity, we refer to core-collapse SNe as SNe in what follows.
Many studies have attempted to use local SNe to get a hint on the dust mass produced per explosion:
mid-infrared observations by \textit{Spitzer} and \textit{AKARI}
have detected dust in some SNe in nearby galaxies. However, the small dust amounts detected
($\sim 10^{-5}$--$10^{-3}$ M$_{\sun}$; \citealt*{kozasa09,gall11}) have proved to be
rather inconclusive due to the difficulty in correctly accounting
for the presence of cold dust in the mid-infrared. \textit{Herschel} observations at far-infrared
(FIR) wavelengths provided an advancement by detecting
a larger amount of dust (0.4--0.7 M$_{\sun}$) in SN 1987A
\citep{matsuura11}.
However, the spatial resolution at FIR wavelengths is not
good enough to isolate the newly formed dust component
from pre-existing dust in
the interstellar medium (ISM). Indeed, an
Atacama Large Millimetre/submillimetre Array (ALMA) observation
of SN 1987A by \citet{indebetouw14} has recently confirmed that the
cold dust component observed by \textit{Herschel} actually
originates in the internal region, that is, the cold dust
detected by \textit{Herschel} is the newly
formed dust component. Resolving the
newly formed SNe dust component in galaxies more distant than
the Magellanic Clouds remains challenging, even with instruments such as ALMA.

Alternatively, the SN production scenario for the first dust
can be tested directly by searching for dust emission from young
high-$z$ galaxies. Galaxies at $z \ga 6$ are favourable for such studies since they
are expected to have minimal contamination from dust formed by old stellar
populations, which typically {could} start dominating the dust budget after
several hundred Myr \citep{valiante09}. Amongst the systematically sampled high-$z$
populations, LAEs are arguably the best dust testbeds since their 
strong Ly$\alpha$ line equivalent width indicates the presence of
a young stellar
population ($\la 200$ Myr, e.g.\ \citealt{finkelstein07,lai07,dayal09}) likely producing the first dust. Indeed, using cosmological simulations
\citet{shimizu10} have shown that the stellar populations in LAEs are mostly younger than
200 Myr at $z>5$, with the contribution from older stellar populations
{becoming} increasingly important at lower redshifts.
Given their young ages, the contribution from low-mass stars such as asymptotic giant branch (AGB) stars
can be neglected \citep{valiante09},
so that the only dust-forming
sources in LAEs can be reasonably assumed to be massive stars (i.e.\ SNe). Moreover, given the typical low metallicities derived for $z\ga 6$ LAEs (0.03--0.3 Z$_{\sun}$;
\citealt*{dayal10}), dust growth in the ISM, which is considered to be the
most effective dust source in nearby galaxies with
nearly solar metallicity \citep{draine09,inoue11,mattsson12},
is most certainly not the dominant process contributing to the total dust budget.
Indeed, \citet{kuo12} have shown that the dominant dust
source switches from SNe dust production to dust
growth at a metallicity of about $0.3$ Z$_{\sun}$. Given the scenario that most dust in LAEs is expected to be produced by SNe, they can be used
to obtain constraints on important quantities such as the amount of dust produced per SNe.

Constraining the dust produced per SN using LAEs {is} challenging
due to the difficulty in directly detecting the 
dust emission from high-$z$ ($z\ga 6$) LAEs, with most observations reporting non-detections using the Plateau de Bure Interferometer
\citep{walter12,kanekar13}. This situation is expected to improve substantially with ALMA: using cosmological
simulations that simultaneously reproduce LAE Ly$\alpha$ and
UV luminosity functions, \citet*{dayal_etal10} predict an average
dust extinction $E(B-V) = 0.15$ \citep[see also][]{nagamine10, kobayashi10}
consistent with values
inferred from spectral energy distribution (SED) fitting
to the rest-frame UV--optical
continuum, $E(B-V)\sim 0.025$--$0.3$
\citep{lai07,pirzkal07,finkelstein09b}. Using these models, \citet{dayal10} argue that ALMA submillimetre
(submm) observations will be able to directly detect the dust emission from LAEs \citep[see also][]{finkelstein09a}.
In contrast, other authors have argued that the extinction of
high-$z$ LAEs
is much lower [$E(B-V)<0.05$] \citep{ouchi08,ono10}.
Directly detecting the dust emission from LAEs
is therefore crucial to quantifying their dust masses and properties, and 
unambiguously identifying SNe as their dominant dust sources and
determining their dust yield. 

Recently \citet[][hereafter O13]{ouchi13} have observed a Ly$\alpha$-emitting
gas blob `Himiko' at $z=6.6$ using ALMA.\footnote{In this paper,
we do not distinguish between Ly$\alpha$ emitters and
Ly$\alpha$ blobs.} This giant LAE was originally discovered as
a \textit{Spitzer} Infrared Array Camera (IRAC) counterpart, 
spectroscopically confirmed  by \citet{ouchi09}. Using ALMA, O13
have put an upper limit of 0.0521~mJy ($3\sigma$) on the observed 1.2 mm flux. 
This low upper limit confirms that
detecting dust in LAEs is difficult probably because they
are experiencing the first dust enrichment. 

Taking advantage of the pioneering observations of Himiko by ALMA, our aim in this paper
is two-fold: firstly, we present a general method to constrain the
dust mass formed per SN that can be directly applied to ALMA observations of young high-$z$ galaxies
such as LAEs. Secondly, we specifically apply our method to
the ALMA observation of Himiko in order to obtain what is arguably the strongest constraint available on the 
dust mass formed per SN using high-$z$ galaxies. To obtain the dust mass per SN, we 
follow the pioneering idea proposed by \citet*{michalowski10b}:
we basically divide the total dust mass by the number
of SNe. They applied the method to
submm-bright galaxies at $z>4$; however, those
galaxies are already evolved with significant dust
enrichment. In such a case, it is difficult to
separate dust formation/growth in SNe, AGB stars, and the ISM.
In fact, \citet{michalowski10b} discuss all those
three formation sources \citep[see also][]{michalowski10a}.
In this paper, we apply their method to LAEs which are at a much earlier 
evolutionary stage as compared to the highly-evolved bright submm galaxies.

The paper is organized as follows: we start by describing
our method to compute the dust
content of a young galaxy in Section~\ref{sec:model}. The dust 
mass so obtained is used to constrain the dust mass produced by a single
SN in Section~\ref{sec:enrichment}. We discuss the results in Section \ref{sec:discussion} and
conclude in Section \ref{sec:conclusion}. We adopt the redshift of Ly$\alpha$ emission
($z=6.595$) for the systemic redshift of Himiko \citep{ouchi09},
and use $(h,\,\Omega_\mathrm{m},\,\Omega_\Lambda)=(0.7,\,0.3,\,0.7)$
for the cosmological parameters.

Finally, we note that using cosmological
galaxy formation simulations, \citet{dayal12} have argued that Lyman break
galaxies (LBGs) and LAEs are derived from the same underlying population of young,
low-metallicity galaxies at $6<z<8$, showing no appreciable differences in terms of their stellar masses, 
ages, metallicities and dust contents. While the exact relation between LAEs and LBGs remains a matter of debate, 
especially {at} $z\ga6$
\citep{verhamme08,nagamine10,dijkstra12}, if the high-$z$ LBG population being observed is indeed as {young and metal-poor as}
LAEs, the method developed in this paper is equally applicable to the latter. 

\section{Constraint on the total dust mass}\label{sec:model}

\subsection{Total dust mass}\label{subsec:Mdust}

Our first aim is to obtain an upper limit on the dust mass
in Himiko, given the ALMA upper-limit of 52.1 $\mu$Jy ($3\sigma$)
at an observational wavelength of 1.2 mm
(a rest wavelength of 158 $\micron$, see O13 for
observational details). While dust is optically thick only
in extreme starbursts such as ultra-luminous infrared
galaxies \citep{soifer99,klaas01,matsushita09},
the dust in LAEs safely satisfies the optically
thin condition in the FIR justifying our use of the optically thin approximation. We now 
explain our formalism for calculating the total dust mass in Himiko. We emphasise that
our formulation itself is general enough to be applied for
future ALMA observations of other high-$z$ galaxies that are optically thin in the FIR.

At FIR wavelengths, dust emission can be treated as
being thermal, so that the flux density
(simply referred to as flux) at an observational frequency
$\nu$ can be written as \citep{dayal10}
\begin{eqnarray}
f_\nu =
\frac{(1+z)\,\kappa_{(1+z)\nu}\, M_\mathrm{d}\, B_{(1+z)\nu}(T_\mathrm{d})}
{d_\mathrm{L}^2},\label{eq:flux}
\end{eqnarray}
where $\kappa_\nu$ is the dust mass absorption coefficient, $M_\mathrm{d}$ and $T_\mathrm{d}$ are
the dust mass and temperature, respectively,
$B_\nu (T)$ is the Planck function, and $d_\mathrm{L}$
is the luminosity distance \citep{carroll92}; given an observed value of $f_\nu$,
equation (\ref{eq:flux}) can be used to obtain $M_\mathrm{d}$
for any given value of $T_\mathrm{d}$.
We express the mass absorption coefficient of dust as
\begin{eqnarray}
\kappa_\nu =\kappa_{158} (\nu /\nu_{158})^\beta ,
\end{eqnarray}
where $\kappa_{158}$ is the value at 158 $\micron$
($\nu_{158}\equiv 1.90\times 10^{12}$ Hz). We take
158 $\micron$ as a reference wavelength since high-$z$
ALMA observations are
often tuned to the rest wavelength of [C\,\textsc{ii}]
emission. \citet{dayal10} adopted mass
absorption coefficients from
\citet{draine84}, referred to as graphite and silicate
in Table \ref{tab:kappa}. Since we expect the
dominant source of dust in LAEs to be SNe, in this work we also use
the mass absorption coefficient for the dust grains formed in SNe. We adopt the
theoretically calculated mass absorption coefficients obtained by
\citet{hirashita05} using the dust species and grain
size distributions in \citet{nozawa03} for dust
condensed in SNe. This case is referred to as SN$_\mathrm{con}$
(Table \ref{tab:kappa}). We also apply the dust
properties after the so-called reverse shock
destruction within the supernova remnant from
\citet{hirashita08}, whose calculation is based on the
dust properties in \citet{nozawa07}. This case is
referred to as SN$_\mathrm{dest}$.
\citet{matsuura11} adopted a mixture of silicate, amorphous
carbon (AC) and iron to estimate the dust mass in SN 1987A
observed by \textit{Herschel},
with the dominant contribution coming from silicate and AC.
With high-resolution ALMA observations,
\citet{indebetouw14} have confirmed that the flux observed by
\textit{Herschel} is actually coming from newly formed dust
in the ejecta in the inner regions.
In order to compare to their estimate of the dust mass,
we also adopt AC (note that their silicate is the same as
we adopted above);
{following \citet{matsuura11}, we use the optical constants
of AC from \citet{zubko96}, and calculate the mass absorption coefficient
by using the Mie theory \citep{bohren83} with the mass density taken from
\citet*{zubko04}.}

\begin{table*}
\centering
\begin{minipage}{100mm}
\caption{Dust properties.}
\label{tab:kappa}
    \begin{tabular}{lccccccc}
     \hline
     Species & $\kappa_{158}\,^\mathrm{a}$ & $\beta\,^\mathrm{b}$ &
     $s\,^\mathrm{c}$
     & $T_\mathrm{eq}\,^\mathrm{d}$ & $M_\mathrm{d}\,^\mathrm{e}$
     & $m_\mathrm{d,SN}\,^\mathrm{f}$ & Ref.\,$^\mathrm{g}$\\
      & (cm$^2$ g$^{-1}$) & & (g cm$^{-3}$) & (K) & ($10^7$ M$_{\sun}$)
     & (M$_{\sun}$)\\
     \hline
     Graphite & 20.9 & 2 & 2.26 & 29.3 & 1.4 & 0.23 & 1, 2\\
     Silicate & 13.2 & 2 & 3.3 & 29.7 & 2.0 & 0.33 & 1, 2\\
     SN$_\mathrm{con}\,^\mathrm{h}$ & 5.57 & 1.6 & 2.96 & 36.7 & 2.7 & 0.45 & 3\\
     SN$_\mathrm{dest}\,^\mathrm{i}$ & 8.94 & 2.1 & 2.48 & 32.9 & 2.2 & 0.37
      & 4\\
     AC$^\mathrm{j}$ & 28.4 & 1.4 & {1.81} & {30.5} & 0.89 & 0.15 & 5, 6\\
     \hline
    \end{tabular}
    
    \medskip

$^\mathrm{a}$Mass absorption coefficient at 158 $\micron$.\\
$^\mathrm{b}$Emissivity index ($\kappa_\nu\propto\nu^\beta$).\\
$^\mathrm{c}$Material density.\\
$^\mathrm{d}$Equilibrium dust temperature in Himiko.\\
$^\mathrm{e}$Upper limits for the dust mass corresponding to the
ALMA $3\sigma$ upper limit.\\
$^\mathrm{f}$Upper limits for the dust mass ejected from a single
supernova (SN).\\
$^\mathrm{g}$ References: 1) \citet{draine84}; 2) \citet{dayal10};
3) \citet{hirashita05}; 4) \citet{hirashita08}; 5) \citet{zubko96};
6) {\citet{zubko04}}.\\
$^\mathrm{h}$Dust condensed in SNe before {reverse} shock destruction.\\
$^\mathrm{i}$Dust ejected from SNe after {reverse} shock destruction.\\
$^\mathrm{j}$Amorphous carbon.
\end{minipage}
\end{table*}

\begin{figure}
\includegraphics[width=0.45\textwidth]{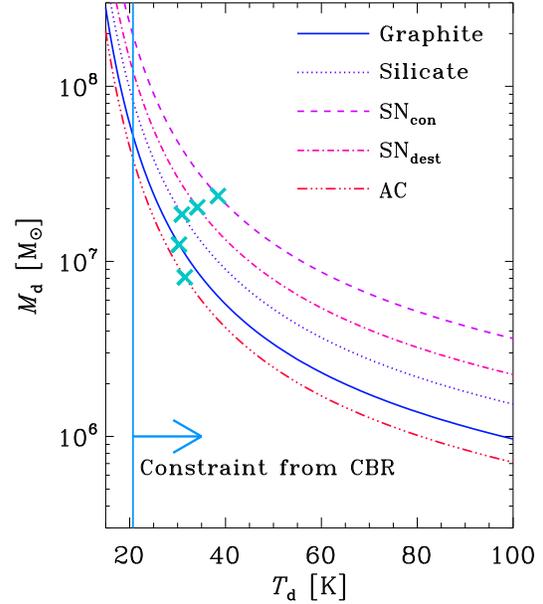}
\caption{Dust mass ($M_\mathrm{d}$) corresponding to
the ALMA $3\sigma$ upper limit for Himiko at $z=6.6$
as a function of the dust temperature ($T_\mathrm{d}$).
The solid, dotted, dashed, dot-dashed, dot-dot-dot-dashed
curves show the dust mass estimated
with the four mass absorption
coefficients in Table \ref{tab:kappa} (graphite,
silicate, SN$_\mathrm{con}$, SN$_\mathrm{dest}$, and
AC, respectively). The cosmic background radiation
(CBR; 20.7 K) shown by the vertical solid line sets
the physical lower limit for $T_\mathrm{d}$.
The cross marks the equilibrium temperature
for each dust species (see Section \ref{subsec:temp}).
\label{fig:Md_Himiko}}
\end{figure}

We show the dust mass
($M_\mathrm{d}$) corresponding to the ALMA $3\sigma$ upper
limit for Himiko as a function of $T_\mathrm{d}$ in
Fig.\ \ref{fig:Md_Himiko}; we use a range of $T_\mathrm{d}$ 
due to the poor constraints available on its value.  
The lower limit of the temperature shown in
Fig.\ \ref{fig:Md_Himiko} (20.7 K) corresponds to the temperature
of the cosmic background radiation at
$z=6.595$; as for the upper limit, we stop at $T_\mathrm{d} = 100$ K
since the dependence of dust mass on the temperature is weak
above this value. Within these temperature limits, we find that
{the} ALMA $3\sigma$ upper
limit corresponds to $M_\mathrm{d}\sim 10^6$--$10^8$ M$_{\sun}$,
depending strongly on the dust temperature. The dust
mass is inversely proportional to the mass absorption
coefficient: $M_\mathrm{dust}$ is smaller by a factor of 5.1
for the AC $\kappa_\nu$ than for the
SN$_\mathrm{con}$ $\kappa_\nu$.

The strong dependence on $T_\mathrm{d}$ makes it
difficult to obtain meaningful constraints on
$M_\mathrm{d}$ unless reasonable temperature estimates are available. Since Himiko has been observed in a single ALMA band, we can not 
constrain $T_\mathrm{d}$ directly from the
SED. We therefore propose an alternative solution to
estimate $T_\mathrm{d}$ in what follows.

\subsection{Dust temperature}\label{subsec:temp}

While {it is} difficult to {constrain the} dust
temperature from the ALMA upper limit,
a reasonable temperature can be obtained 
assuming dust radiative equilibrium, i.e.\ the
equilibrium between dust absorption and emission. This
temperature is referred to as the equilibrium temperature
($T_\mathrm{eq}$) and can be estimated as explained in what follows. 

The total luminosity
emitted by a single dust grain can be written as
\begin{eqnarray}
L_\mathrm{em} & = & \int_0^\infty 4\pi\kappa_\nu
\left(\frac{4}{3}\pi a^3s\right) B_\nu (T_\mathrm{d})\,\mathrm{d}\nu\nonumber\\
 & = & 4\pi\kappa_{158}\nu_{158}^{-\beta}M_\mathrm{d}
 \frac{2h}{c^2}\left(\frac{kT_\mathrm{d}}{h}\right)^{\beta +4}
 \int_0^\infty\frac{x^{\beta +3}}{\mathrm{e}^x-1}\,\mathrm{d}x,
\end{eqnarray}
where $a$ is the grain radius assuming spherical grains, $s$ is the grain material
density (listed in Table \ref{tab:kappa}), $h$ is the Planck constant,
$c$ is the speed of light, $k$ is the Boltzmann
constant and $x = h \nu (kT_\mathrm{d})^{-1}$.
The energy absorbed by a single grain per unit time is estimated as
\begin{eqnarray}
L_\mathrm{abs}=\pi a^2Q_\mathrm{UV}
\frac{L_\mathrm{UV}}{4\pi R^2},\label{eq:abs}
\end{eqnarray}
where $Q_\mathrm{UV}$
is the cross section normalized to the geometric cross
section for the absorption of UV radiation, $L_\mathrm{UV}$
is the total UV luminosity of the stars, and $R$ is the typical
radius of the spatial dust distribution. In this equation,
we assume that the UV radiation dominates the dust
heating \citep[e.g.][]{buat96} and that the UV source is
concentrated in the centre; this latter assumption
does not affect the results in any appreciable manner. For UV radiation, we
assume $Q_\mathrm{UV}=1$ \citep{bohren83,draine84}
since the absorption cross section approaches the
geometrical cross section at wavelengths
as short as 0.2 $\micron$.
The equilibrium temperature, $T_\mathrm{eq}$, can then be
obtained as a solution of
\begin{eqnarray}
L_\mathrm{em}=L_\mathrm{abs}\label{eq:Teq}.
\end{eqnarray}

In order to solve equation (\ref{eq:Teq}), we need to
evaluate $a$, $L_\mathrm{UV}$, and $R$. The values adopted for each of these quantities is now explained: we adopt
a value of $a=0.3~\micron$ since
the dust produced by stars in the early phase of galaxy
evolution is biased towards large grain sizes
($a\sim 0.1$--1 $\micron$;
\citealt{nozawa07,asano13a}) as supported by a number of works:
although \citet{bianchi07} have shown that small grains ($a \ll 0.1~\micron$) are produced
by dust destruction, they do not have a large contribution
to the total dust mass.
\citet{desert90} and \citet{li01} have shown that even if dust grains have a
power-law grain size distribution as in
the ISM of Milky Way, the largest
grains ($a\sim 0.1~\micron$) contribute most to
the dust mass, which is well
traced by the FIR emission. Indeed, \citet{galliano05} have confirmed that the
FIR emission is dominated by big grains
($a\sim 0.1~\micron$) also in nearby
low-metallicity galaxies.
In any case, the equilibrium temperature does
not sensitively depend on $a$
($T_\mathrm{eq}\propto a^{-1/(\beta +4)}$), and the
existence of small grains increases $T_\mathrm{eq}$,
enabling a stricter constraint on the dust mass.
We estimate the total UV luminosity as
$\nu L_{\star\nu}$ ($L_{\star\nu}$ is the stellar
luminosity density at frequency $\nu$) at
rest wavelength 0.2 $\micron$, which is nearly the
centre of the UV wavelength range.
This is equivalent to assuming the
UV wavelength to range between $0.1~\micron$
and $0.3~\micron$, which clearly gives a
conservative (lower) estimate for the equilibrium
dust temperature
since radiation at longer wavelengths also contributes
to dust heating. We also use the
\textit{Hubble Space Telescope}/Wide Field Camera 3
$H_{160}$-band (rest 0.21 $\micron$) flux
for the 0.2 $\micron$ flux of Himiko
(O13), which is converted to
$L_{\star\nu}$. This also gives a conservative (lower)
estimate for the equilibrium dust temperature
because the observed flux is expected be affected by
dust extinction. We obtain
$L_\mathrm{UV}=9.4\times 10^{10}$ L$_{\sun}$ with an
observational error of $\sim 10$ per cent.
Finally, for consistency with O13, $R$ is taken to have a value of 
half of the ALMA beam size, i.e. 0.82 arcsec/2, corresponding to $R=2.2$ kpc
for Himiko.
We note that the dependence of $T_\mathrm{eq}\propto R^{-2/(\beta +4)}$
is weak, and our estimate of $M_\mathrm{d}$ is reasonable as long as dust is distributed within the
ALMA beam; 
indeed, the geometric mean of the semi-major and semi-minor
axis of the stellar distribution is $\simeq 0.4$ arcsec in
Himiko.
Although we cannot completely
exclude the possibility that the dust distribution 
extends much beyond the ALMA beam, a
large amount of dust beyond the stellar distribution
would make it extremely difficult to explain the observed diffuse Ly$\alpha$
emission since it would have been easily attenuated by spatially extended
dust. This possibility of extended
dust distribution needs to be observationally investigated
in the future with more sensitive observations using the full ALMA array; for this paper we
\textit{assume} that the dust distribution is
compact enough, or that all the dust in the galaxy being modelled is contained within a single ALMA beam. The equilibrium dust temperatures
($T_\mathrm{eq}$) calculated for Himiko using these calculations are listed in
Table \ref{tab:kappa} for each dust species.
It is clearly seen that $T_\mathrm{eq}$ tends to be
lower for higher $\kappa_\nu$ in the FIR
because higher $\kappa_\nu$ values indicate more efficient
emission (i.e.\ more efficient radiative cooling of
dust). The exception for this is AC, whose
equilibrium dust temperature is high because of
small $\beta$ and small $s$: small $\beta$ means that the
dust emission at the spectral peak
($\simeq 100~\micron$) is more suppressed due to
inefficient dust cooling and {small $s$} means that
the dust absorbs more UV light resulting in efficient dust heating (see equation \ref{eq:abs}).

It is encouraging to note that our range of $T_\mathrm{eq}$ shown in
Table \ref{tab:kappa} are consistent with observationally
inferred values: \citet{lee12} have inferred dust temperatures
$T_\mathrm{d}\simeq 30$~K for high-$z$ ($z\sim 4$)
LBGs with luminosities comparable to Himiko and \citet{remy13} have found median dust temperatures $T_\mathrm{d}\simeq 32$~K
for a sample of nearby low-metallicity star-forming galaxies, systematically higher than that
in galaxies with solar metallicity. For these temperatures, it is suitable to tune 
ALMA observations to 158 $\micron$
since it is near the emission peak. Indeed, the wavelength
at which the dust emission peaks
($\lambda_\mathrm{peak}$) can be estimated from
the peak of the Planck function as
$\lambda_\mathrm{peak}\simeq 100 (
{T_\mathrm{d}}/{30~\mathrm{K}})^{-1}~\micron$
\citep{rybicki79},
which is close to the [C \textsc{ii}] wavelength
(158 $\micron$), to which high-$z$ observations are often
tuned.

\subsection{Upper limits of the dust mass}
\label{subsec:upper}

Now that we have {estimated} the equilibrium dust temperature
for each dust species in Section \ref{subsec:temp}
(see also Table \ref{tab:kappa}), we can use
equation (\ref{eq:flux}) to obtain the total dust mass 
corresponding to the ALMA $3\sigma$ upper limit flux
for Himiko for each species studied. We use this to put a mark on each line at $T_\mathrm{d}=T_\mathrm{eq}$ in Fig.\ \ref{fig:Md_Himiko}.
The dust mass $M_\mathrm{d}$ at the marked
position on each line then corresponds to the upper limit of the
dust mass in Himiko for a given dust species, and has a value $M_\mathrm{d}<0.89$--$2.7\times 10^7$ M$_{\sun}$; these upper limits for the dust mass are also listed in Table \ref{tab:kappa}.

We remind the reader that it is extremely difficult to put a strong constraint on the dust mass without a constraint on the dust temperature because
low values of $T_\mathrm{d}$ down to the temperature of
the cosmic background radiation
allow dust masses as high as $10^8$ M$_{\sun}$. However,
introducing the physically reasonable assumption
of dust grains being at the equilibrium temperature,
we can put stringent constraints on
{the} dust mass for any given species. Further, the dependence of $T_\mathrm{d}$ on
the mass absorption coefficient makes the dependence
of $M_\mathrm{d}$ on the dust species milder: under
a fixed observed flux,
a smaller $\kappa_\nu$ predicts a larger dust mass,
but it also leads to a higher equilibrium dust temperature because
of less efficient dust FIR emission (i.e.\ dust cooling).
As a consequence, the dust masses obtained under the equilibrium
dust temperature for the various dust species differ only
within a factor of 3, although there is a factor of 5 difference
in the mass absorption coefficient.

\section{Constraint on the dust enrichment}\label{sec:enrichment}

We now investigate the implications of the upper limits on the dust mass derived above. 
There are two major sources of dust 
at $z \ga 6$: dust formation
in SNe, and dust growth by the
accretion of gas-phase metals in the ISM
\citep[e.g.][]{zhukovska08,valiante11,kuo12,asano13b}.
As mentioned before, we have neglected dust enrichment by AGB stars in this
paper, since the age of Himiko is too low
($\la 300$ Myr; O13)
for low-mass stars to have evolved to the AGB phase
\citep{valiante09}. While this argument is true assuming a constant star formation rate which is a good approximation for LAEs (and adopted for Himiko in O13),
\citet{valiante09} have shown that AGB stars may dominate dust production
over SNe as early as 200 Myr, in an environment where all stars form in a burst at $t=0$. However,
bursty star formation histories require
younger ages as compared to a constant star formation history to reproduce the same UV continuum \citep{ono10}. Therefore, we expect that our assumption of neglecting
dust formation in AGB stars is reasonable, regardless of
the star formation history adopted.

\subsection{Dust formation by SNe}\label{subsec:SNe}

Assuming SNe to be the dominant source of dust,
we can constrain the dust mass formed in a single SNe
by using the upper limits derived for the dust mass
in Himiko in Section \ref{subsec:upper}.
The dust mass
ejected from a single SN, $m_\mathrm{d,SN}$ is estimated as
\begin{eqnarray}
m_\mathrm{d,SN}=
\frac{M_\mathrm{d}}{(1-f_\mathrm{dest})N_\mathrm{SN}},
\label{eq:mdSN}
\end{eqnarray}
where $f_\mathrm{dest}$ is the fraction of dust destroyed by
SN shocks in the ISM, and $N_\mathrm{SN}$ is the total
number of SNe. We neglect the effect of dust recycling
in star formation, since the assumption that
SNe are the dominant source over grain growth indicates
an early stage of chemical evolution.
The concept of dust yield per SN (or star) was adopted
by \citet{michalowski10b}, but we include the correction
for the dust destruction by introducing the parameter
$f_\mathrm{dest}$ since dust grains are subject to
destruction in interstellar shocks induced by
the ambient SNe \citep{dwek80,mckee89,jones94,dwek96}.
To solve the above equation, we need to estimate $N_\mathrm{SN}$ and
$f_\mathrm{dest}$, which are obtained as now explained.

The total number of SNe at age $t$, $N_\mathrm{SN}(t)$,
is estimated by
\begin{eqnarray}
N_\mathrm{SN}(t) & = & \int_0^t
\int_{8~\mathrm{M}_\odot}^{40~\mathrm{M}_{\sun}}
\,\psi (t'-\tau_m)\phi (m)\,\mathrm{d}m\,\mathrm{d}t'\nonumber\\
& \simeq & \int_0^t\psi (t')\,\mathrm{d}t'
\int_{8~\mathrm{M}_\odot}^{40~\mathrm{M}_{\sun}}\phi (m)\mathrm{d}m,
\label{eq:N_SN}
\end{eqnarray}
where $\psi (t)$ is the star formation rate at $t$,
$\tau_m$ is the lifetime of a star with mass $m$,
$\phi (m)$ is the IMF, and stars in the mass range of
8--40 $\mathrm{M}_{\sun}$ are assumed to evolve into
SNe \citep{heger03}.
In this paper, the stellar mass, $m$, refers to the mass
at the zero age main sequence. We assume that the lifetimes
of SN progenitors are much shorter than $t$ in order to
simplify the first line of equation (\ref{eq:N_SN}) to the second. 
Further, as in O13, we adopt a Salpeter IMF ($\phi (m)\propto m^{-2.35}$)
between $m_\mathrm{low}=0.1$ and
$m_\mathrm{up}=100$ M$_\odot$. The IMF is normalized
so that the integral of $m\phi (m)$
for the entire mass range is 1.

To link equation (\ref{eq:N_SN}) to observed quantities such as
the stellar mass, we need a few more pieces of information as now explained. The number fraction of stars with mass 
greater than 8 M$_{\sun}$ ($\mathcal{F}_\mathrm{SN}$) can be obtained as 
\begin{eqnarray}
\mathcal{F}_\mathrm{SN}=
\int_{8~\mathrm{M}_\odot}^{40~\mathrm{M}_{\sun}}\phi (m)\mathrm{d}m, 
\end{eqnarray}
where $\mathcal{F}_\mathrm{SN}=6.80\times 10^{-3}$~M$_{\sun}^{-1}$
for the IMF adopted.
As expected, this value is not sensitive to the upper mass used for SNe II; 
changing the upper mass limit to 100 M$_{\sun}$ changes the value slightly to $\mathcal{F}_\mathrm{SN}=7.42\times 10^{-3}$ M$_{\sun}^{-1}$.
Further,
the total stellar mass, $M_*$ can be estimated as
\begin{eqnarray}
M_* & = & \int_0^t\psi (t')\,\mathrm{d}t'\nonumber\\
& & {}-\int_0^t\int_{m_t}^{m_\mathrm{up}}\psi (t'-\tau_m)(m-w_m)\phi (m)\,\mathrm{d}m
\,\mathrm{d}t'\nonumber\\
& \simeq & (1-\mathcal{R})\int_0^t\psi (t')\,\mathrm{d}t' ,
\label{eq:Mstar}
\end{eqnarray}
where $m_t$ is the turn-off mass at age $t$, $w_m$ is
the remnant mass and $\mathcal{R}$ is the returned fraction
of gas from stars, defined as
\begin{eqnarray}
\mathcal{R}=\int_{m_t}^{m_\mathrm{up}}(m-w_m)\phi (m)\,
\mathrm{d}m.
\end{eqnarray}
We have used the instantaneous recycling approximation
\citep{tinsley80} (see also the appendix of
\citealt{hirashita11}) to go from the first to the second line in equation (\ref{eq:Mstar}).
We adopt the turn-off mass at
$t=10^8$ yr (5 M$_{\sun}$) and $w_m$ from
\citet{hirashita11}, obtaining $\mathcal{R}=0.13$.
Combining equations (\ref{eq:N_SN})--(\ref{eq:Mstar}),
we obtain
\begin{eqnarray}
N_\mathrm{SN}\simeq\frac{\mathcal{F}_\mathrm{SN}M_*}{1-\mathcal{R}}.
\label{eq:N_SN_res}
\end{eqnarray}
This equation can be used to obtain the number of SNe
from the total stellar mass.

Using SED fitting O13 inferred the stellar mass of Himiko to be 
$M_*=3.0\times 10^{10}$ and $1.5\times 10^{10}$ M$_{\sun}$, for the case without and with nebular emissions,
respectively.
With $\mathcal{F}_\mathrm{SN}=6.83\times 10^{-3}$ M$_{\sun}^{-1}$
and $\mathcal{R}=0.13$, we obtain
$N_\mathrm{SN}=2.4\times 10^8$ and $1.2\times 10^8$
for the cases without and with nebular emissions.
We hereafter adopt $N_\mathrm{SN}=1.2\times 10^8$ in order to obtain a conservative upper limit
for $m_\mathrm{d,SN}$ for Himiko (see equation \ref{eq:mdSN}).

In order to use equation (\ref{eq:mdSN}), we still need to
estimate $f_\mathrm{dest}$. To this aim,
we utilize the analytical dust
enrichment model presented in \citet{hirashita11} which is now briefly summarized: galaxies are treated as a closed-box single zone and implemented 
with the analytic treatments
adopted in \citet{lisenfeld98} with grain growth
in the ISM added by \citet{hirashita99}.
The model equations are composed of the time evolution of
gas, metals and dust content. While the evolution of gas and metals
is described by recycling in star formation
and production of new metals
in stars, the dust evolution equation
includes formation in stellar ejecta, destruction by
SN shocks and dust mass growth by the accretion of gas-phase
metals in the ISM. After applying the instantaneous recycling
approximation, the equation describing the
relation between dust-to-gas ratio ($\mathcal{D}$) and metallicity
($Z$) can be expressed as
\begin{eqnarray}
\mathcal{Y}\frac{\mathrm{d}\mathcal{D}}{\mathrm{d}Z}=
f_\mathrm{in}(\mathcal{R}Z+\mathcal{Y})-
\left(\beta_\mathrm{SN}+\mathcal{R}-\frac{\beta}{\epsilon}\right)
\mathcal{D},\label{eq:dg_metal}
\end{eqnarray}
where $f_\mathrm{in}$ is the condensation efficiency of
metals in the stellar ejecta (SNe in this paper),
$\mathcal{Y}$ is the mass fraction of newly produced and
ejected metals by stars, $\beta$ is the increment of
dust mass in molecular clouds,
$\epsilon$ is the star formation efficiency in molecular
clouds (assumed to be 0.1)
and
$\beta_\mathrm{SN}$ is the efficiency of dust destruction by
shocks in SN remnants, which is defined by
$\beta_\mathrm{SN}\equiv\epsilon_\mathrm{s}
M_\mathrm{s}\gamma \psi^{-1}$. Here, $\epsilon_\mathrm{s}$
represents the fraction of dust destroyed in a single SN blast,
$M_\mathrm{s}$ is the gas mass swept per SN blast
and $\gamma$ is the SN rate, as introduced in
\citet{dwek80}.
The terms on the right-hand side of equation (\ref{eq:dg_metal})
reflect dust formation in stellar ejecta
[$f_\mathrm{in}(\mathcal{R}Z+\mathcal{Y})$],
dust destruction in SN shocks sweeping the ISM
($-\beta_\mathrm{SN}\mathcal{D}$), dilution of
dust-to-gas ratio by returned gas from stars
($-\mathcal{RD}$), and dust growth in molecular (or dense)
clouds ($\beta\mathcal{D}/\epsilon$).
Except for $\beta$, all the other parameters are taken to
be constants; although $f_\mathrm{in}$
depends on metallicity, such dependence
is predicted to be small (roughly
within a factor of 2) in the metallicity range of
interest \citep{todini01,bianchi07}.
For $\beta$, we adopt the same formula and parameter
values as in the case of silicate in
\citet{hirashita11}.\footnote{Specifically
we adopt Model C in \citet{hirashita11}, i.e.\ a 
power-law grain
size distribution. This distribution is motivated by the fact that
small grain production occurs by shattering at
the metallicity where grain growth becomes efficient
\citep{asano13a}.
In fact, the precise choice of $\beta$ is
not important since LAEs lie in a metallicity range
where dust growth is not important.}
Briefly, $\beta$ is an increasing function of
metallicity, which reflects that dust growth
becomes efficient in a metal-rich condition.

If we only consider dust production by SNe,
$\mathcal{D}=f_\mathrm{in}Z$.
However, if we account for dust destruction by SN shocks,
$\mathcal{D}$ is less than $f_\mathrm{in}Z$, and the
ratio of $\mathcal{D}$ to $f_\mathrm{in}Z$ is related as $(1-f_\mathrm{dest})$. We then use
our model to calculate
$1-f_\mathrm{dest}=\mathcal{D} (f_\mathrm{in}Z)^{-1}$.
We adopt $\mathcal{R}=0.13$ as estimated above.
We evaluate $\mathcal{Y}=0.013$ with the same turn-off mass,
$m_t=5$ M$_{\sun}$.
The dust mass ejected from a single SN, $m_\mathrm{d,SN}$,
is related to $f_\mathrm{in}$ as
\begin{eqnarray}
f_\mathrm{in} & \hspace{-3mm}= & \hspace{-3mm}
\int_{8\,\mathrm{M}_{\sun}}^{40\,\mathrm{M}_{\sun}}
m_\mathrm{d,SN}\phi (m)\,\mathrm{d}m\left/
\int_{8\,\mathrm{M}_{\sun}}^{40\,\mathrm{M}_{\sun}}m_Z(m)\phi (m)
\,\mathrm{d}m\right. ,\nonumber\\
&  &
\end{eqnarray}
where $m_Z(m)$ is the mass of ejected metals from a star
with mass $m$. Adopting a Salpeter IMF, we obtain
$f_\mathrm{in}=0.51(m_\mathrm{d,SN}/\mathrm{M}_{\sun})$.
As shown later in the same section, we obtain $m_\mathrm{d,SN}<0.5$ M$_{\sun}$, resulting in $f_\mathrm{in}<0.3$.

\begin{figure}
\includegraphics[width=0.45\textwidth]{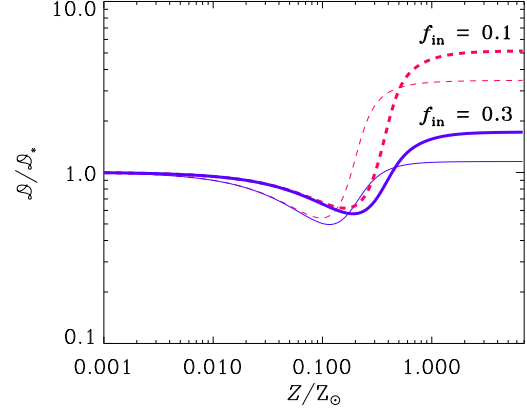}
\caption{
Variation of $\mathcal{D}/\mathcal{D}_*$ ($\mathcal{D}$
is the dust-to-gas ratio and $\mathcal{D}_*$ is the
dust-to-gas ratio expected under the constant
condensation efficiency). We show the cases
with $f_\mathrm{in}=0.3$ and 0.1 by the solid and
dashed lines, respectively. The thick and thin
lines show the cases with $\beta_\mathrm{SN}=9.65$
and 19.3 (normal and enhanced dust destruction), respectively.
The decrease of $\mathcal{D}/\mathcal{D}_*$ is caused
by the dust destruction by SN shocks, while the
increase at high metallicity is due to the grain growth
in the ISM. The minimum of $\mathcal{D}/\mathcal{D}_*$ is
0.57, and 0.62 (0.50 and 0.54) for $f_\mathrm{in}=0.3$ and
0.1, respectively, under $\beta_\mathrm{SN}=9.65$ (19.3).
\label{fig:dust_metal}}
\end{figure}

We now compare the solution of equation (\ref{eq:dg_metal}) with
$\mathcal{D}_*=f_\mathrm{in}Z$, which describes the evolution
of the dust-to-gas ratio for stellar dust production, i.e.\ without
considering dust destruction and growth.
The effects of processes other than stellar dust formation can be seen by dividing the solution of
equation (\ref{eq:dg_metal}) by $\mathcal{D}_*$.
In Fig.\ \ref{fig:dust_metal}, we show the relation between
$\mathcal{D}/\mathcal{D}_*$ and metallicity. As explained before, at low
metallicity, $\mathcal{D}/\mathcal{D}_*\simeq 1$ because
the processes other than dust formation by stars are
negligible. The decrease of $\mathcal{D}/\mathcal{D}_*$
around 0.1 Z$_{\sun}$ is due to dust destruction by SN
shocks in the ISM, after which $\mathcal{D}/\mathcal{D}_*$ increases at $Z\ga 0.2$ Z$_{\sun}$ because of 
grain growth in the ISM. Therefore, while dust can be safely assumed to 
be of stellar origin for $Z\la 0.1$ Z$_{\sun}$, it is impossible to constrain the stellar dust production
for LAEs with metallicity values larger than about $0.2$ Z$_{\sun}$. 
As mentioned at the
beginning of this section, we assume that the dust in LAEs is
predominantly produced by stars (specifically SNe,
because of the young age). In other words, we implicitly assume that the metallicity
of LAEs is lower than
 $\simeq 0.2$ Z$_{\sun}$.

We briefly digress here to discuss the dependence of $\mathcal{D}/\mathcal{D}_*$ on metallicity: the minimum value for this ratio is produced by the combined effects of dust destruction and growth; indeed the depletion of metals in the Milky Way is well reproduced by our models, and determined by the 
balance between dust growth and dust destruction.
The dust-to-gas ratio in other nearby galaxies are also reproduced
by using similar parameters \citep{mattsson14}.
Therefore, our choice of parameters are justified by those data.
However, \citet{hirashita11}
have noted that an enhancement in dust destruction by a
factor of 2 (i.e.\ $\beta_\mathrm{SN}=19.3$) is possible
when considering the uncertainty in the depletion measurements; we therefore also show
cases with $\beta_\mathrm{SN}=19.3$ in Fig.\ \ref{fig:dust_metal}.
In spite of this higher destruction efficiency,
the minimum of $\mathcal{D}/\mathcal{D}_*$ does not
change sensitively. Rather, dust growth enters the
dust evolution earlier (because the gas-phase metals
are more abundant), which indicates that the
metallicity range in which dust growth can be neglected
becomes narrower. Although there is no firm measurement of the metallicity of
Himiko, a clue can be derived from the non-detection
of [C \textsc{ii}] line in O13: these authors have found that [C \textsc{ii}] is weaker by an order of magnitude
with respect to the value expected from the local SFR--[C \textsc{ii}]
relation. Such deficiency of [C \textsc{ii}] emission
has also been seen in local metal-poor galaxies whose
metallicities are below 0.1 Z$_{\sun}$
\citep{delooze14} although its
physical reason has not yet been identified
\citep[see][for possible reasons]{delooze14}. This [C \textsc{ii}] observation provides an empirical
support for the low metallicity of Himiko assumed
in this work and is supported by the work of \citet{vallini13} whose [C \textsc{ii}]
emission models also favour metallicities lower than solar for Himiko. We end by noting the caveat that it is difficult to completely exclude
the possibility of significant contribution from
ISM dust growth to the total dust content. However,
even in this case, the upper limit of $m_\mathrm{d,SN}$
(dust mass produced per SN) is valid, since the excess of dust above that formed in SNe could be attributed to grain growth (or other possible dust sources if any).

We now return to estimating 
$f_\mathrm{dest}$ in equation (\ref{eq:mdSN}).
The fraction of dust destroyed by SN shocks can be
evaluated by the decrement of $\mathcal{D}/\mathcal{D}_*$
in Fig.\ \ref{fig:dust_metal}. The effect of dust
destruction is seen most prominently at the minimum
of $\mathcal{D}/\mathcal{D}_*$ and has a value of 
0.57, 0.62, 0.50, and 0.54 for
$(f_\mathrm{in},\,\beta_\mathrm{SN})=(0.3,\, 9.65)$,
(0.1, 9.65), (0.3, 19.3), and (0.1, 19.3),
respectively. We conservatively adopt a value of 
$1-f_\mathrm{dest}=0.5$ in this work. 

Putting all pieces together, we can now estimate
an upper limit of the dust mass ejected from a single SN,
$m_\mathrm{d,SN}$, for Himiko by using equation (\ref{eq:mdSN}).
The upper limit is listed in Table \ref{tab:kappa} for
each of the dust species adopted. From our calculations, we have successfully obtained upper limits of
$m_\mathrm{d,SN}<0.15$--0.45 M$_{\sun}$ depending on the
dust species; these limits are discussed in the next
section.

\section{Discussion}\label{sec:discussion}

\subsection{Comparison with theoretical dust yields in SNe}

\begin{figure}
\includegraphics[width=0.45\textwidth]{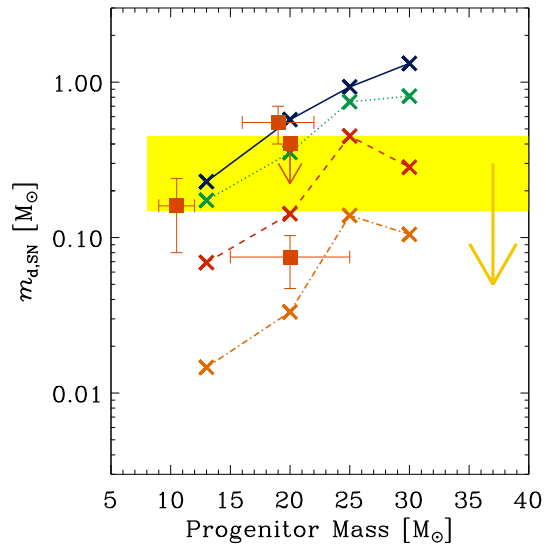}
\caption{The formed dust mass in a supernova (SN) as a function
of progenitor mass at the zero-age main sequence.
The range of upper limits obtained from Himiko is shown by the
shaded area (0.15--0.45 M$_{\sun}$ with the progenitor mass
range appropriate for SNe, 8--40 M$_{\sun}$). The range
corresponds to the different dust species adopted.
The arrow indicates that the shaded area gives upper
limits. The crosses connected by the solid, dotted,
dashed, and dot-dashed lines show the theoretical
predictions of dust condensation and destruction
calculation by \citet{nozawa07}, for various hydrogen
number densities of the ambient medium,
$n_\mathrm{H}=0$, 0.1, 1, and 10 cm$^{-3}$, respectively
(note that $n_\mathrm{H}=0$ corresponds to the
case without reverse shock destruction). To avoid
complication on the diagram, we simply plot the results
for the unmixed helium core in \citet{nozawa07}. Note
that, with the mixed helium core, similar dust masses
are predicted for $n_\mathrm{H}=0$, while the dust tends
to be more destroyed for $n_\mathrm{H}>0$.
\citet{bianchi07} also produce similar dust masses,
but less dependence on the progenitor mass. The
filled squares with error bars show the observed
dust mass detected by \textit{Herschel}
(Table \ref{tab:SN}).
\label{fig:yield}}
\end{figure}

Now that we have obtained constraints on $m_\mathrm{d,SN}$
($m_\mathrm{d,SN}<0.15$--0.45 M$_{\sun}$), we can compare our results 
with theoretically expected values in the literature, as shown in Fig.\ \ref{fig:yield}. We compare to the theoretical data from \citet{nozawa07},
who consider dust condensation in SNe and its
subsequent destruction by reverse shocks
(referred to as reverse shock destruction) under various
hydrogen number densities of the ambient medium
($n_\mathrm{H}$). \citet{bianchi07} have also calculated the dust
condensation and reverse shock destruction in SNe based on
\citet{todini01}, obtaining similar results; therefore the following discussion does not
change even if we adopt the data from \citet{bianchi07}.
Amongst the cases studied by \citep{nozawa07}, we show the case with the
unmixed helium core, but the case with the mixed
helium core predicts similar dust masses with
a tendency towards higher destruction with increasing $n_\mathrm{H}$.

{}From Fig.\ \ref{fig:yield}, it is clear that the
$m_\mathrm{d,SN}$ constraint obtained for Himiko favours the
theoretical dust mass including destruction.
Taking the weighted average of
$m_\mathrm{d,SN}$ for the Salpeter IMF adopted
above, we obtain 0.44, 0.31, 0.13, and 0.036 M$_{\sun}$
for $n_\mathrm{H}=0$, 0.1, 1, and 10 cm$^{-3}$,
respectively ($n_\mathrm{H}=0$ indicates no
reverse shock destruction). Therefore,
the upper limit obtained for Himiko favours
reverse shock destruction with an ambient
density of
$n_\mathrm{H}\ga 1~\mathrm{cm}^{-3}$,
although a scenario with no dust destruction is still permitted
provided we use low-$\kappa_\nu$ dust species such
as SN$_\mathrm{con}$ (Table \ref{tab:kappa}).

\subsection{Comparison with nearby SNe}

In Fig.\ \ref{fig:yield}, we also plot the dust masses
observed in nearby SNe. Although there
are a large number of
SNe observed in {the} mid-infrared by \textit{Spitzer} and
\textit{AKARI}, we limit the sample to the SNe detected
by the \textit{Herschel} far-infrared bands. This is because the mid-infrared observations only trace
the warm dust component, while it is
the cold component traced by FIR observations that
dominates the total newly formed dust budget
\citep{matsuura11}; indeed, mid-infrared observations
obtain dust masses of $\sim 10^{-5}$--$10^{-3}$ M$_{\sun}$
\citep[][and references therein]{gall11},
systematically well below the data from
\textit{Herschel} detections. Further, the newly formed
dust component in SNe should be isolated from the ISM
dust component, which limits the sample to objects
in the Galaxy and Magellanic Clouds.
We therefore use the data for Cas A and Crab for Galactic SN
remnants, and SN 1987A and N 49 for SN remnants in the
Large Magellanic Cloud (LMC) as summarized
in Table \ref{tab:SN}. In fact, although it is difficult
to separate the dust emission from the new and pre-existing components for the SNe in the LMC with the
spatial resolution of \textit{Herschel}, recent observations
of SN 1987A by ALMA have confirmed that the emission comes from the
inner region, i.e. the dust
observed by \textit{Herschel} is indeed the 
newly formed component. Due to possible
contamination, we interpret the dust mass of N 49 to be an upper
limit on the newly formed component.

\begin{table}
\centering
\begin{minipage}{75mm}
\caption{Data of nearby SN remnants
observed by \textit{Herschel}.}
\label{tab:SN}
    \begin{tabular}{lcccc}
     \hline
     Object & $m\,^\mathrm{a}$ & ref$\,^\mathrm{b}$ &
     $m_\mathrm{d,SN}\,^\mathrm{c}$ & ref$\,^\mathrm{d}$\\
      & (M$_{\sun}$) & & (M$_{\sun}$) & \\
     \hline
     Cas A & 15--25 & 1 & $0.075\pm 0.028$ & 1 \\
     Crab  & 9--12  & 2, 3, 4 & 0.08--0.24 & 2 \\
     SN 1987A & 16--22 & 5 & 0.4--0.7      & 3, 4$^\mathrm{e}$ \\
     N 49  & 20     & 6 & $<0.4\,^\mathrm{f}$ & 5\\
     \hline
    \end{tabular}
    
    \medskip

$^\mathrm{a}$Mass of the progenitor at the zero-age main sequence.\\
$^\mathrm{b}$References for the progenitor mass: 1) \citet{krause08};
2) \citet{nomoto82}; 3) \citet{macalpine08}; 4) \citet{gomez12};
5) \citet{arnett89}; 6) \citet{hill95}.\\
$^\mathrm{c}$Inferred dust mass for the newly formed component.\\
$^\mathrm{d}$References for the inferred dust mass:
1) \citet{barlow10}; 2) \citet{gomez12}; 3) \citet{matsuura11};
4) \citet{indebetouw14}; 5) \citet{otsuka10}.\\
$^\mathrm{e}$\citet{indebetouw14} derived a dust mass of
$0.23\pm 0.05$ M$_{\sun}$ by assuming amorphous carbon
while \citet{matsuura11} obtained
$0.35\pm 0.06$ M$_{\sun}$ for the same species. The
difference is explained by the different mass absorption
coefficients adopted in these papers.\\
$^\mathrm{f}$Because of possible contamination with the
molecular cloud with which this SN remnant is interacting,
we regard this dust mass as an upper limit for the newly
formed dust.
\end{minipage}
\end{table}

Comparing the observational data (filled squares)
with the upper {limits} obtained for Himiko
(yellow area) in Fig.\ \ref{fig:yield}, we find
that the data are broadly consistent with our
limits, although the large dust mass derived
for SN 1987A exceeds the upper limits of $m_\mathrm{d,SN}$
obtained for silicate and AC (0.15--0.33 M$_{\sun}$;
Table \ref{tab:kappa}).
\citet{matsuura11} adopted these dust species to
estimate the newly formed dust mass
(0.4--0.7 M$_{\sun}$) in SN 1987A.
We remind the reader that
the dust observed in nearby SNe should suffer
reverse shock destruction before being injected into
the ISM. In other words, the large newly formed dust
mass observed in SN 1987A favours subsequent reverse
shock destruction. 

\subsection{Future prospects for high-redshift
observations}
\label{subsec:future}

In this paper, we have presented the first attempt at
modelling the dust in the high-$z$ ($z\simeq 7$)
Ly$\alpha$-emitting blob Himiko observed by O13.
Further constraints on SNe dust production are
expected from ALMA observations given its increasing capabilities.
The sensitivity converted to the dust mass per
single SN, $m_\mathrm{d,SN}^\mathrm{ALMA}$ can be written as
\begin{eqnarray}
m_\mathrm{d,SN}^\mathrm{ALMA}=\mbox{[(0.15--0.45)~M$_{\sun}$]}
(N/16)^{-1}(t_\mathrm{on}/3.17~\mathrm{h})^{-1/2},
\end{eqnarray}
where $N$ represents the number of antennas and $t_\mathrm{on}$ is 
the on-source integration time; we normalize these quantities using the set-up
in O13. With the planned full ALMA array ($N=66$) becoming operational, we will be able to push $m_\mathrm{d,SN}^\mathrm{ALMA}$
down to $<0.1$ M$_{\sun}$\footnote{http://www.almaobservatory.org/}; such observations will be crucial in
obtaining better constraints on the dust produced per SN.

We emphasize that the method
developed in this paper can be applied
to any high-$z$ object/population wherein SNe are the
dominant dust source. In particular,
based on their cosmological
galaxy formation simulations, \citet{dayal12} have shown that high-$z$ LAEs and LBGs are derived from the same underlying population, with 
LAEs representing a luminous LBG subset. They also show that LBGs
at $z\sim 6$--8 typically have ages $<200$~Myr and
metallicities $\la 0.1$ Z$_{\sun}$. With such young ages and low
metallicities, dust formation by AGB stars and
dust growth in the ISM can safely be neglected in LBGs. In this
context, we can target both LBG and LAE
populations in order to constrain the dust mass produced per SN.

Finally, detection of far-infrared emission lines such as
[C\,\textsc{ii}]~158, [O\,\textsc{ii}] 63, and [N\,\textsc{ii}] 122 $\micron$
can help to constrain ISM physical
properties, especially the metallicity
\citep{nagao12,vallini13}.
Further, CO lines can constrain the gas content, which can
be used in conjunction with the dust content derived from
the continuum to obtain a dust-to-gas ratio.
The relation between dust-to-gas ratio and metallicity
is often used to discuss the dust enrichment history
\citep{lisenfeld98,hirashita99}.
So far, none of the above lines or the continuum have been detected
for Himiko (O13; \citealt{wagg12}); deep ALMA
observations targeting these lines in addition to the
dust continuum will be crucial in constraining the
ISM evolutionary state in Himiko, as well as for
other LAEs at high $z$.

\subsection{Recipe to obtain the dust mass per SN}
\label{subsec:recipe}

For the convenience of future ALMA
observations of LAEs, we summarize our recipe of deriving the dust
mass per SN based on the flux detected; the same method can be used to obtain an upper limit on the dust mass
in case only an observational upper limit is obtained for the flux.
We note that
our analysis is based on the following two assumptions:
[1] the
object has a low metallicity ($Z\la 0.1$~Z$_{\sun}$) so that 
grain growth by accretion of gas-phase metals
can be neglected; and [2] the age is young enough to
neglect the contribution from AGB stars to the total dust
content. We now summarize the main steps in our model:

\noindent
[1] Assuming a dust species, use the corresponding
mass absorption coefficient ($\kappa_\nu$, listed in
Table \ref{tab:kappa}) to estimate the dust temperature, $T_\mathrm{d}$.
If the object is observed at (more than) two rest-FIR wavelengths, it is possible to estimate
$T_\mathrm{d}$ by using the wavelength dependence
of the SED, $\kappa_\nu B_\nu (T_\mathrm{d})$.
However, it is impossible to get the dust temperature
from the SED if the object is observed at a single submm wavelength; in this case, the
dust temperature must be obtained assuming radiative equilibrium (equation \ref{eq:Teq}).

\noindent
[2] Use equation (\ref{eq:flux}) and the same $\kappa_\nu$ as in step [1] above 
to obtain $M_\mathrm{dust}$.

\noindent
[3] Obtain the total stellar mass, $M_*$ by fitting an SED from
a population synthesis code to observations at rest UV wavelengths.
Use this $M_*$ to estimate the total number of SNe
using equation (\ref{eq:N_SN_res}). Using
the same IMF as adopted in this paper
(a Salpeter IMF with the stellar range
0.1--100 M$_{\sun}$), the same values for
$\mathcal{F}_\mathrm{SN}
(=6.80\times 10^{-3}~\mathrm{M}_{\sun}^{-1})$ and
$\mathcal{R}(=0.13)$ can be used. We note that the number of SNe so obtained
is not sensitive to the IMF since both
SNe and rest UV stellar emission reflect the population
of massive stars.

\noindent
[4] Finally, use equation (\ref{eq:mdSN}),
to obtain the dust mass per SN, $m_\mathrm{d,SN}$.
{}The chemical evolution model adopted in
this paper indicates that $f_\mathrm{dest}=0$--0.5
For the purpose of obtaining a conservative upper limit,
$f_\mathrm{dest}=0.5$ is recommended, while adopting
$f_\mathrm{dest}=0$ may lead to an underestimate of
$m_\mathrm{d,SN}$ by a factor of $<2$.

\section{Conclusion}\label{sec:conclusion}

We have examined how the early dust enrichment of
high-$z$ galaxies can be constrained by using
Lyman $\alpha$ emitters (LAEs), which are a young, metal-poor population of galaxies.
Because of their young ages, we assume core-collapse
supernovae (simply referred to as SNe) to be the dominant
source of dust enrichment. Within this scenario, we have provided a `recipe' to constrain
the dust production by SNe in LAEs as summarized
in Section \ref{subsec:recipe}. In particular,
applying our method to Himiko, for which a stringent upper
limit on the dust emission flux (at 1.2 mm) has been obtained by ALMA,
we obtain upper limits on the dust mass produced by a single SN
of 0.15--0.45 M$_{\sun}$; the exact value depends on
the mass absorption coefficient adopted. While these upper limits are
consistent with the values observed in many nearby SN remnants, the 
large dust mass detected in SN 1987A favours
future reverse shock destruction.
Finally, we end by emphasizing that the recipe presented in this work to obtain the
dust mass per SN is applicable to any young, metal-poor galaxy
in which condensation in SN ejecta is the dominant dust formation channel.

\section*{Acknowledgments}

{We are grateful to M. Matsuura and M. Otsuka for providing us with
useful information on dust properties in supernova remnants, and
K. Asada for helpful comments on interferometric observations.
We also thank the anonymous referee for careful reading and
useful comments.}
HH thanks the support from the Ministry of Science and Technology
(MoST) grant 102-2119-M-001-006-MY3. PD acknowledges the support of the European Research Council.



\bsp

\label{lastpage}

\end{document}